\documentclass[showpacs,preprintnumbers,amsmath,amssymb]{revtex4}

\usepackage{graphicx}
\usepackage{bm}
\usepackage{epstopdf}
\usepackage{float}
\usepackage{dcolumn}
\usepackage{bm}
\usepackage{mathrsfs}

\begin{document}

\title{Some exact quasi-normal frequencies of a massless scalar
field in the Schwarzschild space-time}
\author{D. Batic}
\email{davide.batic@ku.ac.ae}
\affiliation{%
Department of Mathematics,\\  Khalifa University of Science and Technology,\\ Main Campus, Abu Dhabi,\\ United Arab Emirates}
\author{M. Nowakowski}
\email{mnowakos@uniandes.edu.co}
\affiliation{
Departamento de Fisica,\\ Universidad de los Andes, Cra.1E
No.18A-10, Bogota, Colombia
}
\author{K. Redway}
\email{kj_redway@hotmail.com }
\affiliation{%
Department of Physics,\\  University of the West Indies,\\ Mona Campus, Kingston,\\ Jamaica} 

\date{\today}

\begin{abstract}
We show that not all quasinormal modes of a massless scalar field in the Schwarzschild metric are encoded in the corresponding characteristic 
equation expressed by means of a continued fraction. We provide an analytical formula for these new hitherto missing quasinormal modes, and in doing that we also construct a generalization of the Gauss convergence criterion.

\noindent

\end{abstract}

\maketitle
\newpage
\pagenumbering{arabic}

\section{Introduction}
For over a century General Relativity continues to fascinate generations of physicists, proposing itself as one of the cornerstones of today's Theoretical Physics. If on the one hand, the underlying equations are formulated in terms of a beautiful simplicity, on the other we find the story of a theory whose progress has been slow over the years due to the complexity of the subject treated. An example is represented by its exact solutions. After the publication of the Einstein field equations in their definitive form in $1915$, Schwarzschild found one year later the expression of the metric in the presence of a massive, static, spherically symmetric, and chargeless body. For the less trivial case of a rotating body we had to wait almost fifty years until Kerr derived the corresponding metric in 1963. Together with the study of exact solutions, General Relativity has brought to  light the presence of objects extraordinarily mysterious as simple: black holes. Only three parameters are needed to explain their structure: mass, charge, and angular momentum. Despite this apparent simplicity, the description of their dynamics is strongly hampered by the manifest non-linearity of Einstein's equations. For this reason their study started only in the 50s with the work of Regge and Wheeler \cite{RG}, in the perturbative framework. If it had been found that a black hole in the presence of a perturbation causes an exponential increase of the latter, then such objects would have been nothing but unstable solutions of the Einstein equations. Therefore, they would be of relative importance at the astrophysical level since they would describe at most a possible transient phase in the life of a star or possibly, of another celestial body. The peculiarity of the perturbative approach is that it allows us to describe the behaviour of black holes in a simple way. In particular, it is found that in the presence of a
generic field the system responds by producing characteristic waves with well-determined frequencies. The dissipation inherent to the event horizon, however, makes the problem non-Hermitian, and therefore, it precludes a standard description of the latter in terms of normal modes. More precisely, the frequencies will present an imaginary part and we will call them quasi-normal modes. Over the years, various useful techniques have been developed for the study of these waves, each with its own peculiarities. For instance, \cite{Ferrari} constructed a quasi-analytic approach to the problem of black-hole oscillations based on a connection between the quasi-normal modes and the bound states of the inverted black-hole effective potential for Schwarzschild, Reissner-Nordstr\"{o}m, and slowly rotating Kerr black holes. Moreover, \cite{IR} used a semi-analytic technique, based on a modified WKB approach, to determine the complex normal-mode frequencies of Schwarzschild black holes. \cite{Leaver} computed the gravitational quasi-normal frequencies of both  stationary and rotating black holes using a continued fraction approach further improved by \cite{Nollert}. Moreover, \cite{Nollert1} showed that the quasi-normal frequencies are the poles of the Green's function for the Laplace-transformed equation for the radial part of scalar as well as electromagnetic or gravitational perturbations of the Schwarzschild space-time: on the basis of this characterization a new technique for the numerical calculation of quasi-normal modes was developed. \cite{Anderson1} investigated the latter problem by constructing a Green's function representation of the solution, which coupled to the semi-analytic phase-integral method allowed to derive approximated formulas determining the excitation of the quasi-normal modes. The role of the highly damped modes and the analysis of the dynamics of the power-law tail for intermediate times in the case of a massless scalar field in the Schwarzschild geometry were addressed in \cite{Anderson2}. Furthermore, \cite{FRAN} took a more operator theoretical approach and developed a new procedure for the computation of the quasi-normal modes by the spectral analysis of the transient scattered waves. Finally, \cite{CAR} studied the power-law tails in the evolution of massless fields around a fixed background geometry corresponding to a Schwarzschild and a Reissner-Nordstr\"{o}m black hole. In the next section we present the relevant equations governing a massless scalar field in the Schwarzschild metric, and give a short derivation of the transcendental equation for the quasi-normal modes involving a continued fraction that was  first obtained in \cite{Leaver}. In Section III we work ''ab origine'', i.e. we analyze the three-terms recurrence relation from which Leaver's  continued fraction stems, by means of the Birkhoff-Adams asymptotic theory for second order linear difference equations coupled to a novel Gauss convergence criterion that we prove at the end of Section III. This allows us to show that not all quasi-normal modes are captured by the aforementioned transcendental equation, and as a result, we obtain a new branch of quasi-normal frequencies for which an exact analytical formula can be provided. This is achieved by allowing one of the coefficients in the recurrence relation to become zero. As a by-product of our method, we are also able to derive the general form of the solution to the three-terms recurrence relation.

\section{The relevant equations}
The manifold of a static black hole of total mass $M$ is described by the Schwarzschild line element ($c=G=1$)
\begin{equation}
ds^2=f(r)dt^2-\frac{dr^2}{f(r)}-r^2(d\vartheta^2+\sin^2{\vartheta}d\varphi^2),\quad f(r):=1-\frac{2M}{r}
\end{equation}
with $r>0$, $0\leq\vartheta\leq\pi$, and $0\leq\varphi<2\pi$. This space-time is spherically symmetric, and therefore, the massless Klein-Gordon equation
\begin{equation}
\square\phi=0,\quad \square:=-\frac{1}{\sqrt{-g}}\partial_\mu(\sqrt{-g}g^{\mu\nu}\partial_\nu\phi) 
\end{equation}
can be separated into spherical harmonics using the ansatz
\begin{equation}
\phi_{\omega\ell m}(t,r,\vartheta,\varphi)=\frac{\psi_{\omega\ell}(r)}{r}Y_{\ell m}(\vartheta,\varphi)e^{-i\omega t} 
\end{equation}
with $\omega$ a complex parameter which may be identified with the energy of the scalar field if $\omega$ is real. Introducing the tortoise coordinate 
\begin{equation}
\frac{dr_*}{dr}=f(r),\quad r_*=r+2M\ln(r-2M),
\end{equation}
the radial equation becomes
\begin{equation}\label{r2}
\frac{d^2}{dr_{*}^2} \psi_{\omega\ell}(r)+\left[\omega^2-V_{eff}(r(r_{*}))\right]\psi_{\omega\ell}(r_{*})=0,\quad r_{*}\in(-\infty,\infty)
\end{equation}
with effective potential
\begin{equation}
V_{eff}(r(r_{*})):=f(r)\left[\frac{1}{r}\frac{df}{dr}+\frac{\ell(\ell+1)}{r^2}\right]=\left(1-\frac{2M}{r}\right)\left[\frac{\ell(\ell+1)}{r^2}+\frac{2M}{r^3}\right],\quad \ell=0,1,2,\cdots,
\end{equation}
where $r$ has to be understood as a function of the Regge-Wheeler coordinate $r_*$. We are interested in the problem of finding those values of the spectral parameter $\omega$ such that equation (\ref{r2}) has non-trivial solutions satisfying quasi-normal modes boundary conditions (QNMBCs). According to \cite{Nollert} the QNMs radial functions must diverge exponentially at the event horizon ($r_*\to-\infty$) and asymptotically far away ($r_{*}\to+\infty)$ while the wave function $\phi$ has an exponential decay in the time variable. This requires that $\omega_I:={\rm{Im}}(\omega)<0$ and the QNMBCs are
\begin{equation}\label{BC1}
\lim_{r_{*}\to-\infty}\psi_{\omega\ell}(r_{*})\approx e^{-i\omega r_{*}},\quad \lim_{r_{*}\to\infty}\psi_{\omega\ell}(r_{*})\approx e^{i\omega r_{*}}.
\end{equation}
It is worth to be mentioned that if we choose instead to write $e^{i\omega t}$ in the separation ansatz, then we need to require that $\omega_I>0$ and the signs appearing in the exponential functions in (\ref{BC1}) will be interchanged. By means of the scaling $x=r/(2M)$ equation (\ref{r2}) can be rewritten as
\begin{equation}\label{r0}
\frac{d}{dx}\left(\frac{x-1}{x}\frac{d\psi_{\omega\ell}}{dx}\right)-\left[\frac{\ell(\ell+1)}{x^2}+\frac{1}{x^3}\right]\psi_{\omega\ell}=-\frac{\Omega^2 x}{x-1}\psi_{\omega\ell},\quad\Omega:=2M\omega,\quad x>1
\end{equation}
and the QNMBCs read
\begin{equation}\label{BC0}
\lim_{x\to 1^+}\psi_{\omega\ell}(x)\approx (x-1)^{-i\Omega},\quad \lim_{x\to+\infty}\psi_{\omega\ell}(x)\approx x^{i\Omega}e^{i\Omega x}.
\end{equation}
It is interesting to observe that equation (\ref{r0}) can be rewritten as a confluent Heun equation in its natural general form \cite{Ronv,Slav}, namely
\begin{equation}\label{r4}
\frac{d^2 \psi_{\omega\ell}}{dx^2}+\left[\sum_{i=1}^2\frac{A_i}{x-x_i}+E_0\right]\frac{d\psi_{\omega\ell}}{dx} +\left[\sum_{i=1}^2\frac{C_i}{x-x_i}+\sum_{i=1}^2\frac{B_i}{(x-x_i)^2}+D_0\right]\psi_{\omega\ell}=0,\quad x_1=0,\quad x_2=1, 
\end{equation}
with
\begin{equation}
A_2=B_1=-A_1=1,\quad B_2=D_0=\Omega^2,\quad C_1=\ell(\ell+1)+1,\quad C_2=2B_2-C_1,\quad E_0=0.
\end{equation}
Furthermore, our $(\ref{r0})$ and $(\ref{BC0})$ coincide with (3) and (4) in \cite{Leaver} for $\epsilon=-1$, $2M=1$, and $\rho=-i\Omega$. A solution to (\ref{r0}) exhibiting the desired behaviour at the event horizon and at space-like infinity may  be represented by the Ansatz \cite{Jaffe,Baber}
\begin{equation}\label{ansatz}
\psi_{\omega\ell}(x)=(x-1)^{\rho}x^{-2\rho}e^{-\rho(x-1)}f_{\omega\ell}(x),\quad f_{\omega\ell}(x)=\sum_{n=0}^{\infty}a_n\left(\frac{x-1}{x}\right)^n,\quad \rho:=-i\Omega,
\end{equation}
where $f_{\omega\ell}$ must satisfy the following differential equation
\begin{equation}\label{CHII}
\frac{d^2 f_{\omega\ell}}{dx^2}+\left[1-c+\frac{1-2c}{x}+\frac{c}{x-1}\right]\frac{df_{\omega\ell}}{dx}+\left[\frac{d}{x}-\frac{d}{x-1}+\frac{c^2}{x^2}\right]f_{\omega\ell}=0,\quad x>1
\end{equation}
with 
\begin{equation}\label{cd}
c=1+2\rho=1-4iM\omega,\quad d=c^2+(c-1)^2+\ell(\ell+1).
\end{equation}
Introducing the new coordinate $y=(x-1)/x$ the ODE (\ref{CHII}) becomes
\begin{equation}\label{odef}
y(y-1)^2\frac{d^2 f_{\omega\ell}}{dy^2}+\left[(2c+1)y^2-4cy+c\right]\frac{df_{\omega\ell}}{dy}+(c^2 y-d)f_{\omega\ell}=0,\quad 0<y<1,\quad f_{\omega\ell}(y)=\sum_{n=0}^{\infty}a_n y^n,
\end{equation}
and it is not difficult to verify that in agreement with (6), (7), and (8) in \cite{Leaver}, the coefficients $a_n$ satisfy the three-terms recurrence relation
\begin{eqnarray}
&&ca_1-da_0=0,\quad a_0=1,\label{rr1}\\
&&\alpha_n a_{n+1}+\beta_n a_n+\gamma_n a_{n-1}=0,\quad n\geq 1\label{rr2}
\end{eqnarray}
with 
\begin{equation}\label{rr3}
\alpha_n=(n+1)(c+n),\quad \beta_n=-[d+2n(2c+n-1)],\quad \gamma_n=(c+n-1)^2.
\end{equation}

\section{Analysis of the recurrence relation}
The strategy followed in \cite{Nollert,Leaver} to compute the quasi-normal frequencies consists mainly of two steps. One first expresses the solution of the radial part of the Klein-Gordon equation (\ref{r0}) subject to the boundary conditions (\ref{BC0}) 
as a power series of the form (\ref{ansatz}), where the coefficients of the power series expansion in (\ref{ansatz}) satisfy the three-terms recurrence relation represented by (\ref{rr1}) and (\ref{rr2}). Then, one determines the QNMs by some numerical/analytical method. More precisely, in the seminal work of \cite{Leaver} QNMs were found by computing the zeroes of a certain continued fraction associated to the recurrence relation (\ref{rr1}) and (\ref{rr2}). No matter which method is used in the second part, it is clear that in the first step the spectral parameter $\omega$ is a QNM if the following conditions are met, namely
\begin{enumerate}
\item
the boundary conditions (\ref{BC0}) at the even horizon and at infinity are fulfilled.
\item
The recurrence relation (\ref{rr2}) has a minimal (subdominant, recessive) solution, i.e. a  solution $\varphi_n$ of (\ref{rr2}) such that $\varphi_n/x_n\to 0$ as $n\to\infty$ for any solution $x_n$ of (\ref{rr2}) that is not a multiple of $\varphi_n$ \cite{Elaydi}. If this is the case, then the convergence of the power series (\ref{ansatz}) at infinity can be checked by the Gauss criterion \cite{Knopp} stating that the infinite series $\sum_{n=0}^\infty a_n$ converges if the ratio $a_{n+1}/a_n$ can be written in the form
\begin{equation}\label{GC}
\frac{a_{n+1}}{a_n}=1-\frac{\alpha}{n}-\frac{\theta_n}{n^\lambda},\quad\alpha,\lambda>1
\end{equation}
with $\theta_n$ a bounded sequence, i.e. there exists some positive constant $K$ such $|\theta_n|<K$ for every nonnegative integer $n$. For cases where the Gauss criterion is not applicable we develop a suitable generalization proved at the end of this section.
\item
After substitution of the candidate QNM into (\ref{rr1}) and in (\ref{rr2}) the initial conditions should be satisfied and the recurrence relation itself should not give rise to an under/overdetermined system of equations for the coefficients. 
\end{enumerate}
\cite{Leaver} showed that the convergence problem arising from condition 2. above can be translated into a convergence problem for a certain continued fraction that can be used to set up a transcendental equation for the quasi-normal frequencies. Let us recall the procedure used in \cite{Leaver} to obtain the continued fraction. If we set $y_{n+1}:=a_{n+1}/a_n$, the difference equation (\ref{rr3}) can be written as $y_n=-\gamma_n/(\beta_n+\alpha_n y_{n+1})$ from which we obtain the representation
\begin{equation}\label{cont_frac}
\frac{a_n}{a_{n-1}}=-\cfrac{\gamma_n}{\beta_n-\cfrac{\alpha_n\gamma_{n+1}}{\beta_{n+1}-\cfrac{\alpha_{n+1}\gamma_{n+2}}
{\beta_{n+2}-\cfrac{\alpha_{n+2}\gamma_{n+3}}{\beta_{n+3}-\ddots}}}} .
\end{equation}
Finally, if we impose the initial condition $\alpha_0 a_1+\beta_0 a_0=0$, and we make use of  (\ref{cont_frac}), we get the following transcendental equation
\begin{equation}\label{QNMs_frac}
\beta_0 -\overset{\infty}{\underset{n=1}{\text{K}}}\left(\frac{\alpha_{n-1}\gamma_n}{\beta_n}\right)=0,
\end{equation}
where K is the standard notation for a continued fraction. The above expression is meaningless as long as the continued fraction is not convergent. In order to ensure convergence, it is necessary to require $\gamma_n\neq 0$ and the existence of a minimal solution for the recurrence relation (\ref{rr3}) \cite{Elaydi}. Note that the condition $\gamma_n\neq 0$ is equivalent to the requirement that $\alpha_{n-1}\neq 0$ for all $n$ because $\gamma_n=\alpha^2_{n-1}/n^2$ as it can be readily checked from (\ref{rr3}). 

We outline a mathematical tool which we will use later to analyze the recurrence relation. Since our recurrence relation (\ref{rr2}) is a special case of the more general recurrence relation
\begin{equation}\label{yrr}
y_{n+2}+p_1(n)y_{n+1}+p_2(n)y_n=0
\end{equation}
with coefficients $p_i$ admitting asymptotic representations 
\begin{equation}\label{coff}
p_1(n)=\sum_{s=0}^\infty\frac{\mathfrak{a}_s}{n^s},\quad
p_2(n)=\sum_{s=0}^\infty\frac{\mathfrak{b}_s}{n^s},\quad\mathfrak{b}_0\neq 0,
\end{equation}
the existence and construction of a minimal solution can be readily obtained by means of the so-called Birkhoff-Adams asymptotic theory for second order linear difference equations of the form (\ref{yrr}) and (\ref{coff}) \cite{Elaydi,Galbrun,B1,A1,B2,Wong}. According to this theory, asymptotic solutions to (\ref{yrr}) are classified in terms of the roots of the characteristic equation $\lambda^2+\mathfrak{a}_0\lambda+\mathfrak{b}_0=0$. In particular, if $\lambda_1=\lambda_2$, which is the relevant case for (\ref{rr2}), and $2\mathfrak{b}_1\neq\mathfrak{a}_0\mathfrak{a}_1$, then there exist two linearly independent solutions to (\ref{yrr}) having asymptotic representations
\begin{equation}\label{sns}
y_{\pm,n}=\lambda^n e^{\pm\gamma\sqrt{n}}n^\alpha\sum_{s=0}^\infty\frac{\mathfrak{c}_{\pm,s}}{n^{s/2}}
\end{equation}
with
\begin{eqnarray}
\alpha&=&\frac{1}{4}+\frac{\mathfrak{b}_1}{2\mathfrak{b}_0},\quad
\gamma=2\sqrt{\frac{\mathfrak{a}_0\mathfrak{a}_1-2\mathfrak{b}_1}{2\mathfrak{b}_0}},\label{alphgamm}\\
\mathfrak{c}_{\pm,1}&=&\frac{1}{24\mathfrak{b}_0^2\gamma}\left[\mathfrak{a}_0^2\mathfrak{a}_1^2-9\mathfrak{b}_0^2-32\mathfrak{b}_1^2+8\mathfrak{a}_0\left(\mathfrak{a}_1\mathfrak{b}_1-3\mathfrak{a}_2\mathfrak{b}_0\right)+24\mathfrak{b}_0\left(\mathfrak{b}_1+2\mathfrak{b}_2-\mathfrak{a}_0\mathfrak{a}_2\right)\right].\label{cfrakpm}
\end{eqnarray}
Casting the recurrence relation (\ref{rr2}) into the form  $a_{n+2}+p_1(n)a_{n+1}+p_2(n)a_n=0$ with
\begin{equation}\label{RRn2}
p_1(n)=-\frac{d+2(n+1)(2c+n)}{(n+2)(c+n+1)},\quad p_2(n)=\frac{(c+n)^2}{(n+2)(c+n+1)}.
\end{equation}
and observing that 
\begin{equation}
p_1(n)=-2+\frac{4-2c}{n}+\frac{2c^2+2c-d-8}{n^2}+\mathcal{O}\left(\frac{1}{n^3}\right),\quad
p_2(n)=1+\frac{c-3}{n}+\frac{7-2c}{n^2}+\mathcal{O}\left(\frac{1}{n^3}\right),
\end{equation}
then, (\ref{sns}) together with (\ref{alphgamm}) and (\ref{cfrakpm}) leads immediately to the following asymptotic solutions
\begin{equation}\label{hass}
a_{n,\pm}=e^{\pm\gamma\sqrt{n}}n^\alpha\sum_{j=0}^\infty\frac{\mathfrak{c}_{\pm,j}}{n^{j/2}},\quad
\alpha=\frac{2c-5}{4},\quad\gamma=2\sqrt{c-1},\quad \mathfrak{c}_{\pm,0}=1,\quad
\mathfrak{c}_{\pm,1}=\pm\frac{112c^2-104c-48d+31}{48\sqrt{c-1}}.
\end{equation}
Note that $\Re{\gamma}>0$ for $\Im{\Omega}<0$, and $\Re{\Omega}<0$, while the real part of $\gamma$ is negative if $\Im{\Omega}<0$, and $\Re{\Omega}>0$. This means that in both cases there are always one exponentially increasing and one exponentially decreasing solution. The solution characterized by the exponential fall off is the minimal solution for which (\ref{QNMs_frac}) is convergent. This minimal solution combined with the Gauss criterion is also needed to show the convergence of the power series solution (\ref{ansatz}) at infinity. In particular, it turns out that the condition $\Im{\omega}<0$ is a necessary and sufficient condition to ensure the existence of a minimal solution. Furthermore, the study of the existence of a minimal solution allows to reduce the region of the complex plane where candidate QNMs can be found.

In the reminder of this section we show that the transcendental equation (\ref{QNMs_frac}) does not capture all quasi-normal frequencies. This is achieved by deriving a new branch of QNMs  thanks an educated guess suggested by the recurrence relation (\ref{rr2}). More precisely, we investigate those spectral families in the parameter $\omega$ such that one of the coefficients in (\ref{rr2}) vanishes for a certain choice of a positive integer. As a quite a surprise, it turns out that it is not only possible to obtain values of $\omega$ that satisfy all the three conditions introduced at the beginning of this section, and hence, they can be identified as genuine QNMs but also a relatively simple new analytic formula describing them can be provided. To warm up, let us consider the case when the coefficient $\gamma_n$ in (\ref{rr2}) becomes zero for a certain choice of a positive integer. Assume that $c_j=1-j$ with $j=1,2,\cdots$ where $j\neq 0$ because otherwise $\Re{(\Omega)}=0=\Im{(\Omega)}$, and fix any positive integer $j$. Then, using (\ref{rr3}) yields $\gamma_{n,j}=(n-j)^2$ signalizing that this coefficient vanishes once when $n=j$. For this choice of $c_j$, we  get a purely imaginary family in the spectral parameter given by  
\begin{equation}\label{QNMSf}
\omega_j=-i\kappa\frac{j}{2},\quad j=1,2,\cdots,\quad\kappa=\frac{1}{2M},
\end{equation}
where $\kappa$ denotes the surface gravity of the black hole. It is not difficult to verify for these values of the spectral parameter that the recurrence relation (\ref{rr2}) admits always a minimal solution ensuring convergence at space-like infinity through the Gauss criterion. To this purpose, let us cast (\ref{rr2}) into the form (\ref{yrr}) so that the Birkhoff-Adams asymptotic theory can be applied. In this case, $p_1$ and $p_2$ admit asymptotic expansions with coefficients
\begin{equation}\label{koeff}
\mathfrak{a}_0=-2,\quad \mathfrak{a}_1=2(j+1),\quad \mathfrak{a}_2=-[4j+5+\ell(\ell+1)],\quad
\mathfrak{b}_0=1,\quad \mathfrak{b}_1=-(j+2),\quad \mathfrak{b}_2=2j+5,
\end{equation}
and there will be two linearly independent solutions with asymptotic representations 
\begin{equation}\label{caso1}
a_{\pm,n}\sim \frac{e^{\pm 2i\sqrt{jn}}}{n^{\frac{3}{4}+\frac{j}{2}}}\sum_{s=0}^\infty\frac{\mathfrak{c}_{\pm,s}}{n^{s/2}},\quad
\mathfrak{c}_{\pm,0}=1,\quad \mathfrak{c}_{\pm,1}=\mp\frac{i}{\sqrt{j}}\left[\frac{j^2}{3}-\frac{j}{2}-\frac{3}{16}-\ell(\ell+1)\right].
\end{equation}
Furthermore, using (\ref{caso1}) together with the fact that $\Re{(c_{\pm,1})}$ vanishes yields 
\begin{equation}\label{ratioasympt}
\left|\frac{a_{\pm,n+1}}{a_{\pm,n}}\right|=\left[1-\frac{2j+3}{4n}+\mathcal{O}\left(\frac{1}{n^2}\right)\right]
\left[1-\frac{\Re{(c_{\pm,1})}}{2n^{3/2}}+\mathcal{O}\left(\frac{1}{n^2}\right)\right]=1-\frac{\sigma}{n}+\mathcal{O}\left(\frac{1}{n^2}\right),\quad\sigma:=(2j+3)/4.
\end{equation}
Since the quotient $|a_{\pm,n+1}/a_{\pm,n}|$ has an asymptotic expansion of the form (\ref{ratioasympt}) with $\sigma>1$, then the Gauss criterion (see p. 280 in \cite{Knopp}) implies that $\sum_{n=0}^\infty|a_{\pm,n}|$ is convergent, and therefore, $\sum_{n=0}^\infty a_{\pm,n}$ is absolutely convergent. This ensures that the series in (\ref{ansatz}) converges at space-like infinity. However, (\ref{QNMSf}) cannot represent a new branch of QNMs because it fails to satisfy the initial condition (\ref{rr1}) of the recurrence relation. This can be easily seen in the following cases. Let us write explicitly the first equations for the coefficients emerging from (\ref{rr1}) and (\ref{rr2}). For $c_j=1-j$ with $j=1,2,\cdots$ we find
\begin{eqnarray}
&&(1-j)a_{1,j}-d_j a_0=0,\quad a_0=1,~d_j=(j-1)^2+j^2+\ell(\ell+1),\label{a1}\\
&&\alpha_{1,j}a_{2,j}+\beta_{1,j}a_{1,j}+\gamma_{1,j}a_0=0,\label{a2}\\
&&\alpha_{2,j}a_{3,j}+\beta_{2,j}a_{2,j}+\gamma_{2,j}a_{1,j}=0,\label{a3}
\end{eqnarray}
and so on. For instance, in the case $j=0$ equation (\ref{a1}) becomes $0\cdot a_{1,1}-[1+\ell(\ell+1)]a_0=0$ which violates the initial condition in (\ref{rr1}). At this point one could be tempted to change initial conditions and try for instance, $a_0=0$ and $a_1=1$. Then, (\ref{a2}) becomes $2a_{2,1}-[1+\ell(\ell+1)]\cdot 1=0$ which can be solved for $a_{2,1}$ and so on. The problem with these modified initial conditions is that they will ensure convergence asymptotically at infinity but the radial function will be identically zero at the event horizon, and therefore, the first constraint in (\ref{BC0}) will be violated. If we consider the case $j=2$, (\ref{a1}) and (\ref{a2}) become $-a_{1,2}-[5+\ell(\ell+1)]a_0=0$ and $0\cdot a_{2,2}-[1+\ell(\ell+1)]a_{1,2}+a_0=0$. Clearly, this system can never be fulfilled. The same problem occurs for other choices of $j$. Following a similar argument, one may set $c_j=-j$ with $j$ nonnegative integer so that the first coefficient in the recurrence relation (\ref{rr2}) vanishes once for $n=j$. Also in this case no new QNMs emerge.

We come now to the core of our results. A new branch of quasi-normal frequencies satisfying all requirements can be obtained by imposing that the coefficient $\beta_n$ in (\ref{rr2}) vanishes for some choice of a nonnegative integer. Starting with a fixed nonnegative integer $j$ we set 
\begin{equation}\label{cpm}
c_{\pm,j}=-j+\frac{1}{2}\pm\frac{i}{2}\sqrt{1+2\ell(\ell+1)}.
\end{equation}
Then, the coefficient $\beta_n$ in (\ref{rr2}) can be rewritten as 
\begin{equation}\label{betan}
\beta_{n,j,\pm}=2(j-n)\left(j-n\pm i\sqrt{1+2\ell(\ell+1)}\right),
\end{equation}
and it clearly vanishes for $n=j$. For $c$ chosen according to (\ref{cpm}) we find the following spectral family
\begin{equation}\label{qnmsII}
\widehat{\omega}_{\pm,j\ell}=\pm\frac{\kappa}{4}\sqrt{1+2\ell(\ell+1)}-i\frac{\kappa}{2}\left(j+\frac{1}{2}\right),\quad j=0,1,\cdots,
\end{equation} 
where for instance, $\widehat{\omega}_{+,j\ell}$ is obtained from $c_{-,j}$ using (\ref{cd}). We need to check if the frequencies given by (\ref{qnmsII}) can be interpreted as QNMs. We first verify that (\ref{rr2}) admits a minimal solution. Substituting (\ref{cpm}) into (\ref{rr2}) and bringing (\ref{rr2}) into the form of the recurrence relation (\ref{yrr}), the coefficients $p_1$ and $p_2$ in (\ref{yrr}) have asymptotic expansions of the form (\ref{coff}) with expansion coefficients
\begin{eqnarray*}
\mathfrak{a}_{0,\pm}&=&-2,\quad 
\mathfrak{a}_{1,\pm}=2j+3\mp i\sqrt{1+2\ell(\ell+1)},\quad \mathfrak{a}_{2,\pm}=-4j-7-\ell(\ell+1)\pm 2i\sqrt{1+2\ell(\ell+1)},\\
\mathfrak{b}_{0,\pm}&=&1,\quad \mathfrak{b}_{1,\pm}=-j-\frac{5}{2}\pm\frac{i}{2}\sqrt{1+2\ell(\ell+1)},\quad 
\mathfrak{b}_{2,\pm}=2j+6\mp i\sqrt{1+2\ell(\ell+1)}.
\end{eqnarray*} 
Using (\ref{sns}), (\ref{alphgamm}), and (\ref{cfrakpm}) two linearly independent solutions can be constructed having asymptotic representations
\begin{equation}\label{aeaf}
a_{\pm,n}\sim n^{\alpha_\pm} e^{\gamma_\pm\sqrt{n}}\sum_{s=0}^\infty\frac{\mathfrak{c}_{\pm,s}}{n^{s/2}},
\end{equation}
where
\begin{eqnarray}
\alpha_{\pm}&=&-\left(1+\frac{j}{2}\right)\pm\frac{i}{4}\sqrt{1+2\ell(\ell+1)},\quad
\gamma_\pm=\sqrt{-4j-2\pm 2i\sqrt{1+2\ell(\ell+1)}},\\
\mathfrak{c}_{\pm,0}&=&1,\quad \mathfrak{c}_{\pm,1}=\frac{1}{24\gamma_\pm}\left[\frac{2}{3}j^2\mp\frac{1}{3}\left(1+2i\sqrt{1+2\ell(\ell+1)}\right)j-\frac{7}{8}-\frac{7}{3}\ell(\ell+1)\pm\frac{i}{6}\sqrt{1+2\ell(\ell+1)}\right].
\end{eqnarray}
We recall that $a_{-,n}$ stems from the choice $c_{-,j}$. In the following it is useful to rewrite $\gamma_\pm$ as
\begin{equation}\label{ag}
\gamma_\pm=2\sqrt[4]{1+(2j+1)^2+2\ell(\ell+1)}\left(\pm\sin{\frac{\alpha}{2}}+i\cos{\frac{\alpha}{2}}\right),\quad \alpha=\arctan{\frac{\sqrt{1+2\ell(\ell+1)}}{2j+1}}\in(0,\pi/2).
\end{equation}
By means of the asymptotic expansion (\ref{aeaf}) we find that
\begin{equation}\label{ratio}
\left|\frac{a_{\pm,n+1}}{a_{\pm,n}}\right|=1\pm\frac{K}{2\sqrt{n}}+\mathcal{O}\left(\frac{1}{n}\right),\quad K:=\left|\Re{(\gamma_{\pm})}\right|=2\sqrt[4]{1+(2j+1)^2+2\ell(\ell+1)}\sin{\frac{\alpha}{2}}
\end{equation}
with $\alpha$ as in (\ref{ag}). At this point a couple of remarks are in order. The case with the plus sign in the above expression can be disregarded because it does not ensure convergence at space like infinity. Hence, the only relevant case to be considered is the one for $a_{-,n}$. Furthermore, the presence of the $n^{-1/2}$ term in the asymptotic expansion (\ref{ratio}) signalizes that the Gauss criterion (\ref{GC}) cannot be applied in its present form. However, a generalization of the aforementioned criterion is possible. To this purpose, let $\mathfrak{A}_n:=|a_{-,n}|$. Then, (\ref{ratio}) can be rewritten as 
\begin{equation}
\frac{\mathfrak{A}_{n+1}}{\mathfrak{A}_n}=1-\frac{K}{2\sqrt{n}}+\mathcal{O}\left(\frac{1}{n}\right).
\end{equation}
Clearly, for some positive integer $N$ we have for all $n\geq N$
\begin{equation}
\frac{\mathfrak{A}_{n+1}}{\mathfrak{A}_n}\leq 1-\frac{K}{2\sqrt{n}},
\end{equation}
or equivalently,
\begin{equation}\label{ungl1}
\sqrt{n}\mathfrak{A}_{n+1}\leq(\sqrt{n}-1)\mathfrak{A}_n-\beta\mathfrak{A}_n,\quad\beta:=\frac{K}{2}-1>0,
\end{equation}
where the second inequality above can be expressed as 
\begin{equation}\label{ineqjl}
\sqrt[4]{1+(2j+1)^2+2\ell(\ell+1)}\sin{\frac{\alpha}{2}}>1
\end{equation} 
with $\alpha$ given in (\ref{ag}). On the other hand, we can rewrite (\ref{ungl1}) as
\begin{equation}
(\sqrt{n}-1)\mathfrak{A}_n-\sqrt{n}\mathfrak{A}_{n+1}\geq\beta \mathfrak{A}_n>0
\end{equation}
but $\sqrt{n-1}\geq\sqrt{n}-1$ for all $n\geq 1$ and therefore, we can conclude that
\begin{equation}\label{ungl2}
\sqrt{n-1}\mathfrak{A}_n-\sqrt{n}\mathfrak{A}_{n+1}\geq\beta\mathfrak{A}_n>0.
\end{equation}
This shows that for all $n\geq N$ the sequence $(\mathfrak{x})_{n\geq N}$ with $\mathfrak{x}_n=\sqrt{n}\mathfrak{A}_{n+1}$ is a monotonously decreasing sequence. Since the terms of this sequence are positive, the sequence itself converges to some limit value $\xi\geq 0$. At this point, it is convenient to rewrite (\ref{ungl2}) as
\begin{equation}
\mathfrak{A}_n\leq\frac{\mathfrak{d_n}}{\beta},\quad\mathfrak{d}_n:=\mathfrak{x}_{n-1}-\mathfrak{x}_n.
\end{equation}
Since the series $\sum_n\mathfrak{d}_n$ converges, we can immediately conclude that $\sum_n\mathfrak{A}_n$ converges as well. Therefore, $\sum_{n} a_{-,n}$ is also convergent, and the power series in (\ref{ansatz}) with $a_n=a_{-,n}$ will also converge at space-like infinity. The result of this analysis shows that only the values 
\begin{equation}\label{qnm2}
\widehat{\omega}_{+,j\ell}=\frac{\kappa}{4}\sqrt{1+2\ell(\ell+1)}-i\frac{\kappa}{2}\left(j+\frac{1}{2}\right) 
\end{equation}
of the spectral parameter lead to the existence of a minimal solution to (\ref{rr2}) provided that the condition (\ref{ineqjl}) is satisfied. Because of this constraint not all choices for $j$ and $\ell$ in (\ref{qnm2}) are admissible. For instance, the inequality (\ref{ineqjl}) is satisfied for $\ell\geq 2$ when $j=0$, $\ell\geq 3$ when $j=1,2$, for $\ell\geq 4$ when $j=3,4$, and so on. Furthermore, we need to verify that after substitution of $c_{-,j}$  into (\ref{rr1}) and (\ref{rr2}) the initial conditions are satisfied and the recurrence relation itself does not give rise to an under/overdetermined system of equations for the coefficients. Let us write explicitly the first equations for the coefficients emerging from (\ref{rr1}) and (\ref{rr2}). For $c_{-,j}$ with $j=0,1,2,\cdots$ and given by (\ref{cpm}) we find
\begin{eqnarray}
&&c_{-,j}a_{1,-,j}-d_{-,j} a_0=0,\quad a_0=1,\label{aa1}\\
&&\alpha_{1,-,j}a_{2,-,j}+\beta_{1,-,j}a_{1,-,j}+\gamma_{1,-,j}a_0=0,\label{aa2}\\
&&\alpha_{2,-,j}a_{3,-,j}+\beta_{2,-,j}a_{2,-,j}+\gamma_{2,-,j}a_{1,-,j}=0,\label{aa3}
\end{eqnarray}
and so on. For instance, in the case $j=0$ we have $d_{-,0}=0$ and since $c_{-,j}$ does not vanish for any value of $j$ we find from (\ref{aa1}) that $a_{1,-,0}=0$. Note that since $c_{-,j}$ is complex none of the coefficients $\alpha_{n,-,j}$ and $\gamma_{n,-,j}$ in (\ref{rr2}) can be zero for some choice of $j$. Observe that from (\ref{aa2}) we obtain $a_{2,-,0}=-\gamma_{1,-,0}/\alpha_{1,-,0}$. Moreover, (\ref{aa3}) allows to compute $a_{3,-,0}$ and so on. In the case, $j=1$ we have $d_{-,1}\neq 0$ and $\beta_{1,-,1}=0$. It is straightforward to verify that we can recursively obtain all the unknown coefficients. The same applies also to other choices of $j$. This shows that (\ref{qnm2}) represents a new branch of QNMs.

We conclude this section by deriving the general form of the solution of the recurrence relation (\ref{rr2}). Let $y_n:=|a_n|$. Then, using (\ref{hass}) yields
\begin{equation}
\lim_{n\to\infty}\frac{y_{n+1}}{y_n}=\lim_{n\to\infty}\left|\frac{a_{n+1}}{a_n}\right|=1.
\end{equation}
Choose $z_n:=\ln{y_n}$. Then, $z_{n+1}-z_n\to 0$ as $n\to\infty$. Hence, for a given $\epsilon>0$ we can find a positive integer $N$ such that $|z_{n+1}-z_n|<\epsilon/2$ for all $n\geq N$. Furthermore, for $n\geq N$ we have
\begin{equation}
|z_n-z_N|\leq\sum_{s=N+1}^n|z_s-z_{s-1}|<\frac{\epsilon}{2}(n-N)
\end{equation}
and therefore,
\begin{equation}
\left|\frac{z_n}{n}\right|<\frac{\epsilon}{2}\left(1-\frac{N}{n}\right)+\frac{|z_N|}{n}<\frac{\epsilon}{2}+\frac{\epsilon}{2}=\epsilon
\end{equation}
for $n$ large enough. We can conclude that $z_n/n\to 0$ as $n\to\infty$, or equivalently $z_n=n\nu_n$ for some null sequence $\nu_n$. This implies that the general representation for the solution of (\ref{rr2}) is $a_n=\pm e^{n\nu_n}$.

\section{Conclusions}
By analyzing the recurrence relation for the solution of a massless particle in a Schwarzschild metric we found a new branch of previously missing quasi-normal modes, see equation (\ref{qnm2}). The frequencies in the second branch can be interpreted as resonances, i.e., states with a definite energy and life-time. We have proved meticulously the QNMs found fulfil all conditions imposed on the frequencies. Furthermore, we also constructed the general representation for the solution of the recurrence relation (\ref{rr2}). It is interesting to observe that the radial parts of the massless/massive Klein-Gordon and Dirac equations in the presence of a Schwarzschild, Reissner-Nordstr\"{o}m, or Kerr-Newman black hole are all represented by second-order linear differential equations with rational coefficients, and thus they will give rise to three-terms recurrence relations that are amenable to be treated by our method. This will be treated ina  future work.

\acknowledgments
We thank the Referee for her/his interest in our work and for helpful comments that greatly improved the manuscript.

\end{document}